\documentclass[12pt]{article}
\usepackage{mymacros}

\title{A multiverse model in $T^2$ dS wedge holography}
\author[a]{Sergio E. Aguilar-Gutierrez,}
\author[b]{Filip Landgren}
\affiliation[a]{Institute for Theoretical Physics, KU Leuven, Celestijnenlaan 200D, B-3001 Leuven, Belgium}
\affiliation[b]{Mathematical Sciences and STAG Research Centre, University of Southampton, Highfield, Southampton SO17 1BJ, UK}
\emailAdd{sergio.ernesto.aguilar@gmail.com, f.landgren@soton.ac.uk}
\abstract{We construct a multiverse model where empty AdS$_{d+1}$ space is cut off by a pair of accelerated dS$_d$ space universes, at a finite AdS boundary cutoff {which we treat as a $T^2$ deformation in the holographic dual}, and one in the AdS interior, the IR brane{; and denote the construction as $T^2$ dS wedge holography}. We glue together several copies of this configuration along the UV cutoff and the IR branes in a periodic matter. To provide the model with dynamics similar to those of near Nariai black holes used in other multiverse toy models{, we specialize to $d=2$ and} add dS JT gravity as an intrinsic gravity theory on the IR branes. We then study the entanglement entropy with respect to a finite cutoff observer, who finds a Page curve transition due to an entanglement island connecting the UV cutoff and IR brane. This process involves the coarse-graining of information outside the causally accessible region to the observer. Our model provides an explicit realization of entanglement between IR and UV degrees of freedom encoded in the multiverse.}

\begin{document}

\maketitle

\section{Introduction}\label{sec:intro}
Recently, there has been a lot of attention on double holographic models with the development of wedge holography \cite{Akal:2020wfl,Miao:2020oey,Miao:2021ual}, where gravity on a (1+d)-dimensional anti-de Sitter (AdS$_{d+1}$) space region bounded by a pair of end-of-the-world (ETW) branes is dual to a CFT$_{d - 1}$ theory living on the interception between the branes. This is realized within the Karch-Randall (KR) braneworld models \cite{Randall:1999ee,Randall:1999vf,Karch:2000gx,Karch:2000ct,Giddings:2000mu}.

Perhaps one of the most exciting prospects in this program is to learn lessons that can be applied to spacetimes relevant to cosmology. The KR-type models have allowed the development of different cosmological models \cite{Cooper:2018cmb,Fan:2021eee,Antonini:2019qkt,VanRaamsdonk:2021qgv,Waddell:2022fbn,Antonini:2022blk,Antonini:2022xzo,Antonini:2022opp,Antonini:2022fna}, as well as several applications for studying quantum information observables with braneworld holography \cite{Almheiri:2019hni,Chen:2020uac,Chen:2020jvn,Chen:2020hmv,Emparan:2020znc,Cotler:2022weg,Emparan:2022ijy,Grimaldi:2022suv,Aguilar-Gutierrez:2023tic,Chen:2023tpi,Geng:2019ruz, Geng:2020kxh, Geng:2020qvw, Geng:2020fxl, Geng:2021wcq, Geng:2021iyq, Geng:2021hlu, Geng:2021mic, Geng:2022slq, Geng:2022tfc, Geng:2022dua, Geng:2023iqd} and formal aspects of higher curvature gravities \cite{Hu:2022lxl,Miao:2023mui,Aguilar-Gutierrez:2023kfn}.

Since the ETW branes can have arbitrary cosmological constants, this has led to different realizations of wedge holography, and importantly for us, to the development of de Sitter (dS) wedge holography \cite{Aguilar-Gutierrez:2023tic}. Here, one employs a dS$_d$ space ETW branes near the asymptotic boundary of AdS$_{d+1}$ space and in the interior bulk geometry, denoted as the ultraviolet (UV) and infrared (IR) branes respectively.\footnote{There are also notions of flat/dS space wedge holography; see \cite{Ogawa:2022fhy,Bhattacharjee:2022pcb} for recent developments.} {By placing the UV brane sufficiently close to the boundary to effectively render the gravity non-dynamical, \cite{Aguilar-Gutierrez:2023tic} studied {the holographic entanglement entropy \cite{Ryu:2006bv,Ryu:2006ef,Hubeny:2007xt,Faulkner:2013ana,Engelhardt:2014gca} and holographic complexity \cite{Susskind:2014rva,Stanford:2014jda,Belin:2021bga, Belin:2022xmt}} with respect to an observer living in the UV cutoff universe to be studied\footnote{Other approaches to quantum information on dS braneworlds can be found in \cite{Geng:2021wcq,Yadav:2023qfg,Cotler:2022weg}.}.}
In particular, for AdS$_3$ ambient space, one can analytically reproduce a Page curve with respect to a UV observer. Given that pure Einstein gravity is topological in 3-dimensions, one normally either introduces fluctuations in the braneworld location \cite{Geng:2022slq,Geng:2022tfc}, or intrinsic gravity theories on the brane. The latter approach has allowed progress in different areas, such as for providing new hints in the context of the information paradox \cite{Chen:2020uac,Chen:2020hmv,Miao:2023unv,Li:2023fly}, and developing bounds the intrinsic gravity couplings on the ETW branes based on consistency with entanglement velocity \cite{Lee:2022efh}. The higher curvature corrections to the intrinsic gravity on the brane render the gravity dynamical, as opposed to purely topological. This allows for a more consistent holographic treatment of branes with intrinsic gravity and sharp formulations of quantum information observables \cite{Lee:2022efh}. It has been found that introducing the intrinsic gravitational theory reproduces a Page curve with massless gravitons in AdS wedge holography \cite{Miao:2023unv}, and cone holography \cite{Li:2023fly}.

Our work aims at exploring the coarse-graining of information encountered in semi-classical quantum cosmology \cite{Hartle:2016tpo,Aguilar-Gutierrez:2021bns} {by proposing an extension to dS wedge holography, where instead of a second ETW brane near the asymptotic AdS boundary, we locate a finite boundary cutoff generated by a $T^2$ deformation \cite{Hartman:2018tkw} in the dual CFT, which is kept very close to the asymptotic AdS boundary} { - we refer to this as $T^2$ dS wedge holography}. We perform this step in order to impose Dirichlet boundary conditions at a finite radial location\footnote{{ Different works of wedge holography with non-trivial observables consider that at least one of the ETW branes is connected with a non-gravitating bath, such as a boundary CFT \cite{Caceres:2021fuw,Geng:2021mic,Ghosh:2021axl,Caceres:2020jcn,Anderson:2020vwi,Bhattacharya:2021jrn,Geng:2021iyq,Qi:2021sxb}. 
{ Here, we will instead treat the regulated UV brane as a $T^2$ deformation, with a well-defined Dirichlet boundary at a finite fixed radial distance \cite{Zamolodchikov:2004ce, Taylor:2018xcy, Hartman:2018tkw}}}, and evaluate quasi-local diffeomorphism invariant observables; where the closeness of the finite cutoff region to the asymptotic boundary allows us to neglect the effect of non-localities, at scales of the deformation parameter \cite{Cardy:2019qao}, {in the dual QFT \cite{Apolo:2023ckr, Banerjee:2024wtl,Demise:2021cfx,Chen:2018eqk,Donnelly:2018bef}}.} In the context of eternal inflation in quantum cosmology, such as false vacuum eternal inflation\footnote{False vacuum decay refers to a tunneling event from a local minimum in an inflaton potential to the global minimum. The reader is referred to \cite{Coleman:1980aw} for a pioneering publication in vacuum decay in semiclassical gravity, and \cite{Garriga:2006hw} for original work on false vacuum eternal inflation.}, one expects that meta-observers in a given universe have access to the information available in other universes given a large amount of redundancy in the theory \cite{Hartle:2016tpo,Aguilar-Gutierrez:2021bns}, and yet they cannot interact with those (spacelike separated) universes. This coarse-grained information shares similarities with the fine-grained von Neumann entropy computed with entanglement islands \cite{Penington:2019npb,Almheiri:2019psf}, as both are determined from the saddle points from the semi-classical gravitational path integral i.e. a coarse-graining over the allowed geometrical configurations of the theory. 

\begin{figure}[t!]
\centering
    \includegraphics[width=0.75\textwidth]{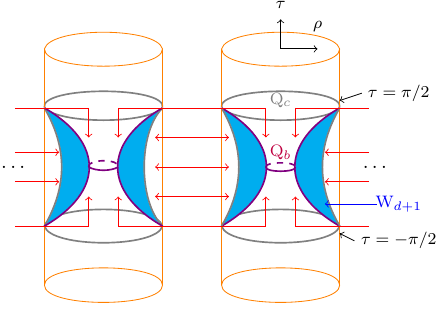}
\caption{``dS wedge holographic" multiverse model formed with a pair of dS$_d$ branes, labeled by Q$_b$ (in purple) and Q$_c$ (gray) embedded in AdS$_{d+1}$ space (orange), and ending on a pair of codimension-two Euclidean defects at global AdS time $\tau=\pm\frac{\pi}{2}$. The region between the branes (blue) is denoted as the wedge W$_{d+1}$. We glue together different copies of the same system along the UV cutoff and IR brane (represented by the red arrows) to form a double-sided brane wedge configuration for both the UV cutoff and IR brane. From the brane perspective, there is a junction between dS$_d$ universes on opposite sides. The respective Penrose diagram is shown in Fig. \ref{fig:repManifold}.}
\label{fig:UV IR branes}
\end{figure}

In this work, we study this situation within a multiverse dS-braneworld toy model, to gain access to holographic tools that allow us to evaluate the holographic entanglement entropy of a subregion in a multiverse through the Hubeny-Rangamani-Takayanagi (HRT) formula \cite{Ryu:2006bv,Ryu:2006ef,Hubeny:2007xt}. This will help us interpret the information a UV observer has access to in terms of CFT degrees of freedom. 
To carry out this task, we introduce the braneworld model illustrated in Fig \ref{fig:UV IR branes}.\footnote{Throughout the letter, we employ latin indices for braneworld coordinates.}
This model can be viewed as an extension of dS wedge holography \cite{Aguilar-Gutierrez:2023tic}. A technical difference with \cite{Aguilar-Gutierrez:2023tic} is that besides connecting different bulk spacetimes through the dS branes, we also add an intrinsic gravity theory in the IR ETW branes; {besides generating the UV cutoff through a $T^2$ deformation in the dual CFT, as mentioned above.} This allows for the effective braneworld description to contain dynamics even for ($2+1$)-dimensional bulk space with a dS$_2$ space brane in the IR region of the bulk and a UV cutoff. We adopt dS \emph{Jackiw-Teitelboim} (JT) gravity \cite{JACKIW1985343, TEITELBOIM198341} \footnote{This theory can describe the perturbations around the near horizon region of arbitrary dimensional near extremal Schwarzschild-de Sitter (SdS) black holes, known as the near Nariai limit, or from the dimensional reduction of pure dS$_3$ space. We will consider only the near Nariai perspective. See, for instance, \cite{Svesko:2022txo} and references therein for general aspects of these theories.} as the intrinsic theory on the IR branes, while keeping the UV cutoff. This is achieved by imposing Dirichlet (Neumann) boundary conditions on the UV (IR), and it allows us to fix the cosmological constant in both the finite cutoff region and IR brane to be dS space. Having intrinsic gravity on the brane modifies the gravitational observables anchored to the UV boundary, such that we can study the consequences for the properties in the multiverse model.  

Our main point of interest is the evolution of holographic entanglement entropy with respect to {the $T^2$-deformed CFT degrees of freedom} in the UV cutoff within this model.\footnote{See \cite{Levine:2022wos,Yadav:2023qfg,Aguilar-Gutierrez:2021bns,Pasquarella:2022ibb,Jiang:2024xnd,Yadav:2024ray} for closely related previous studies.} {On the technical level, in order to have well-defined notion of holographic entanglement entropy, we need to restrict the analysis to a quasi-local description of the dual field theory (see e.g. \cite{Apolo:2023ckr, Banerjee:2024wtl}), which is realized when the length of the entangling region in the HRT formula is much larger than the length scale set by the deformation parameter.} 
Under this restriction for the entangling region, our construction has an apparent black hole-like information loss paradox. Based on \cite{Aguilar-Gutierrez:2023tic}, we consider a UV observer collecting Hawking radiation. 
 The von Neumann entropy with respect to the UV cutoff observer describes two phases; at early times, it is a monotonically increasing function of time during the so-called Hawking phase. The observer could access more Hawking modes than the total number of degrees of freedom on the IR brane union with the entangling region in question. However, later there is a phase where it decreases and then saturates, which is an example of the entanglement islands \cite{Penington:2019npb,Almheiri:2019psf}. { In the connected phase, we find that the entanglement entropy is non-vanishing and depends on degrees of freedom in both the IR and the UV; the minimal surface after the Page transition is anchored to both of the branes.}  

Our results suggest that the information available to each observer living in a given UV cutoff completely encodes that of the other universes, as a consequence of the Israel junction conditions \cite{Israel:1966rt}  in the configuration (see Sec. \ref{sec:coarsegraing multiverse}). This observation confirms that the braneworld multiverse model explicitly realizes the coarse-graining of information for observables in quantum cosmology \cite{Hartle:2016tpo,Aguilar-Gutierrez:2021bns}. Moreover, the island configurations we encounter show that the different observers cannot transmit nor decode messages between universes. This result is consistent with the central dogma \cite{Shaghoulian:2021cef} and the no-cloning theorem for cosmological horizons discussed in \cite{Levine:2022wos}. The central dogma in this context refers to encoding information of the spacetime beyond the cosmological horizon, with respect to a given observer, from the interior region. In \cite{Levine:2022wos}, it was noticed in a multiverse toy model that observers in spacelike separated regions could in principle encode regions of spacetime with some overlap, such that they could reconstruct information from those regions without affecting the ability of the other observer to do the same. This would then enter into tension with the no-cloning theorem. However, the authors also find additional non-overlapping island saddles with respect to meta-observers on a non-dynamical gravity region. The dominance of these saddles implies that the apparent paradox simply does not arise. The model in \cite{Levine:2022wos} has many similarities to ours. The non-dynamical region in our setting refers instead to the UV cutoff where one collects Hawking radiation. The island transitions show that the RT surfaces will be confined to a single dS wedge universe. The main new observation, explicit in our model, is the entanglement between the UV and IR degrees of freedom present in the multiverse model, and in the coarse-graining of information with respect to different UV observers, who take the role of the meta-observers in the false vacuum eternal inflation models.

{Lastly, our proposal for a system with a UV and IR cutoff region also relates to recent works where double-trace $T\overline{T}$ deformations in a CFT dual to pure AdS$_3$ space with an ETW brane allow for a new realization of the Page transition \cite{Deng:2023pjs}}. In contrast to this other approach, the dS cutoff regions do not overlap with each other, except at particular time slices (see Fig. \ref{fig:UV IR branes}) which modifies considerably the analysis, with consequences for the Page curve transition, and the quantum cosmology interpretation, which we persuade in Sec. \ref{sec:coarsegraing multiverse}.

\textbf{Outline}: The rest of this paper is organized as follows. We start by reviewing the geometric construction and we introduce our multiverse braneworld model in Sec. \ref{sec: model review}. In Sec. \ref{sec:coarsegraing multiverse}, we present an information recovery protocol and compute the holographic entanglement entropy of spacetime subregions using the HRT formula, which results in a page curve transition with respect to a UV observer. In Sec. \ref{sec:central dogma} we comment on the connection between our model with the central dogma and non-cloning theorem in the context of quantum cosmology. We conclude in Sec. \ref{sec:discussion} with a discussion of our findings and important questions to be addressed in the future. 

\section{Braneworld (multiverse) model}\label{sec: model review}
To formulate the multiverse model, we start constructing its building block as a single AdS$_{d+1}$ bulk geometry with an ETW brane inside the bulk, denoted the IR brane, and the finite cutoff (produced by a $T^2$ deformation in the dual theory). This simple model is a { modification} of dS wedge holography that appeared in \cite{Aguilar-Gutierrez:2023tic}. {Importantly, instead of considering an ETW brane very close to the asymptotic boundary,
 we will consider a $T^2$ deformation of the original CFT$_d$ theory living on a dS$_{d}$ background dual to pure AdS$_{d+1}$ with the appropriate foliation. This allows for consistency at the moment of imposing Dirichlet boundary conditions at a finite boundary location, which will allow us to find a Page curve transition in the multiverse toy model in Sec. \ref{sec:coarsegraing multiverse}.}. In this section, we {specify the geometric construction,} including an IR brane without intrinsic gravity, and later, we introduce multiple ETW branes with an intrinsic dS JT gravity theory on them.

\subsection{De Sitter wedge holography, and \texorpdfstring{$T^2$}{} deformations}
The original proposal in dS wedge holography \cite{Aguilar-Gutierrez:2023tic} considered a double holographic AdS$_{d+1}$ space bounded by a single pair of dS$_d$ space near the asymptotic boundary (denoted as the UV region, where gravity decouples) and an arbitrary location in the bulk interior (the IR region). {The configuration was interpreted as the Hartle-Hawking preparation of state corresponds to a tunneling instanton describing membrane creation \cite{Aguilar-Gutierrez:2023tic} in the IR. {Our configuration, however, considers an important modification, where the finite cutoff UV region is instead due to a $T^2$ deformation in the dual CFT (see Sec. \ref{ssec:T^2}).}}

There are 3 equivalent ways to describe the system (see Fig. \ref{fig:UV IR branes}):
\begin{itemize}
    \item[a.] A pair of codimension-two Euclidean conformal defects on S$^1\times$S$^{d-2}$-spaces that are timelike separated from each other.
    \item[b.] A pair of entangled dS$_{d}$ universes with CFT$_{d}$ matter connected during the infinite past and future via transparent boundary conditions.
    \item [c.] AdS$_{d+1}$ bulk space with a pair of dS$_{d}$ cutoff region, where the IR corresponds to a Randall-Sundrum brane \cite{Randall:1999ee}\footnote{As noticed in an alternative Randall-Sundrum multiverse model \cite{Yadav:2023qfg}, the dS branes would have a finite lifetime. In our configuration the accelerating branes produce a singularity at the location where the branes nucleate \cite{Garriga:1993fh,Arcos:2022icf} (a big bang), as well as where they decay (a big crunch) \cite{Emparan:2022ijy}, where the lack of unitary on the codimension-two dS conformal defects is manifest.} that overlap at global AdS time $\tau=\pm\frac{\pi}{2}$, while the UV region corresponds to a $T^2$ deformation \cite{Hartman:2018tkw} in the dual CFT$_d$.
\end{itemize}

The system is described by the following action
\begin{equation}\label{eq:Ihigherdim}
\begin{aligned}
    I=&\int_{\hat{\mathcal{M}}}\rmd^{d+1} x\sqrt{-\hat{g}}\qty[\frac{1}{16\pi G_{d+1}}\qty(\hat{R}-2\Lambda_{d+1})+\hat{\mathcal{L}}_{\rm bulk}]+\frac{1}{8\pi G_{d+1}}\int_{\partial \hat{\mathcal{M}}}\rmd^d x\sqrt{-h}\,K\\
    &+\sum_{i=b,~c}\qty(\int_{Q_i}\rmd^d x\sqrt{-h^{(i)}}\mathcal{L}^{(i)}_{\rm intrinsic}+I^{(i)}_{\rm matter})\,,
    \end{aligned}
\end{equation}
where
\begin{equation}
    \Lambda_{d+1}=-\frac{d(d-1)}{2\ell_{d+1}^2}\,.
\end{equation}
In the above $h_{\mu\nu}$ is the induced metric on $\partial \mathcal{M}$; $\hat{\mathcal{L}}_{\rm bulk}$ is the bulk field Lagrangian density; $Q_c$ is the UV cutoff region, and $Q_b$ denote (the worldvolume) of the IR ETW brane, where we consider both regions to be double sided; $\mathcal{L}^{(i)}_{\rm intrinsic}$ is the intrinsic gravity theory on the $Q_i$ regions; $h^{(i)}_{ij}$ the respective induced metric (where $i,\,j$ are $2$-dimensional indices); $I^{(b)}_{\rm matter}$ is the matter field theory on $Q_i$; and $G_{d+1}$ is Newton's gravitational constant.

We describe this spacetime with AdS global coordinates
\begin{equation}\label{eq:global coordinates}
\rmd s^2=\rmd \rho^2-\cosh^2\rho\,\rmd\tau^2+\sinh^2\rho(\rmd\alpha^2+\cos^2\alpha\,\rmd\Omega_{d-2}^2)~.
\end{equation}
In this foliation, dS branes of arbitrary tension can only be found in the range $-\frac{\pi}{2}\leq\tau\leq\frac{\pi}{2}$ \cite{Parikh:2012kg}.\footnote{The UV cutoff region and the IR ETW brane overlap at global AdS time $\tau=\pm\pi/2$ as seen from (\ref{eq:map global to dS foliation}) for $t\rightarrow\pm\infty$ and $\rho\rightarrow\infty$ in both regions.}

We can also employ a change of coordinates from global AdS$_{d+1}$ space to AdS$_{d+1}$ with dS$_d$ space foliations to place the ETW branes, which will have an effective positive cosmological. In general, the metric has the form \cite{Arcos:2022icf},
\begin{equation}
    \rmd s^2=\ell_{d+1}^2\qty[H^2\sinh^2\sigma\,\rmd s_{\text{dS}}^2+\rmd\sigma^2]~.\label{eq:main metric}
\end{equation}
where $\rmd s_{\rm dS}$ is a $d$-dimensional line element for dS space in any coordinate system, and $H$ is the Hubble rate. In these coordinates, we locate the finite IR and UV cutoff regions at $\sigma=\sigma_b$ and $\sigma=\sigma_c$ respectively, such that 
\begin{equation}
W_{d+1}:\quad \sigma_b<\sigma<\sigma_c    ~.
\end{equation}
To describe the evolution with respect to the global time of an observer living in the dS finite UV cutoff, it is most convenient to use a Rindler-AdS$_{d+1}$ background with global coordinate dS$_d$ foliation with the explicit mapping
\begin{equation}
\begin{aligned}\label{eq:map global to dS foliation}
    \tan\tau&=\sinh t\tanh\sigma~,\\
\sinh{\rho}&=\sinh{\sigma}\cosh t~,
\end{aligned}
\end{equation}
and recover the metric
    \begin{equation}\label{eq: AdS with global dS foliation}
\rmd s^2=\ell_{d+1}^2\qty[\rmd \sigma^2+H^2{\sinh^2\sigma}(-\rmd t^2+\cosh^2t~(\rmd\alpha+\cos^2\alpha\,\rmd\Omega_{d-2}^2)]~.
\end{equation}
From now on we use a rescaling of coordinates where $\ell_{d+1}=1$ and $H=1$.

\subsection{{Incorporating \texorpdfstring{$T^2$}{} deformations}}\label{ssec:T^2}
$T^2$ deformations were defined by \cite{Hartman:2018tkw}\footnote{See also \cite{Taylor:2018xcy,Morone:2024ffm}.} as a generalization of the finite cutoff interpretation of $\text{T}\overline{\text{T}}$ for CFT$_d$ in arbitrary dimensions. The definition is based on the $\text{AdS}_{d+1}/\text{CFT}_{d}$ dictionary relating the bulk gravity and CFT partition functions, which is taken to hold at finite bulk radial cutoff $r_{B}$
\begin{equation}
    Z_{\text{EFT}}[r_{B};\gamma_{ij},J]=Z_{\text{grav}}[h_{ij}^{B}=r_{B}^{2}\gamma_{ij},\psi_{B}=r_{B}^{\Delta-d}J]\;.\label{eq:Dirchdic}
\end{equation}
The left-hand side is the generating function for the (assumed holographic) effective field theory, which need not be a CFT itself, $h_{ij}$ is the metric describing the field theory geometry, and $J$ is, for simplicity, taken to be a source for a scalar operator $\mathcal{O}$ of dimension $\Delta$. On the right-hand side is the (on-shell) gravitational partition function in an asymptotically AdS background with metric 
\begin{equation}\label{eq:asymptotic AdS metric}
\rmd s^{2}=g_{\mu\nu}\rmd x^{\mu}\rmd x^{\nu}=\frac{\rmd r^{2}}{N(r)}+r^{2}\gamma_{ij}\rmd x^{i}\rmd x^{j}~,
\end{equation}
where $N(r)\to r^{2}$ near the conformal boundary; the bulk metric and bulk scalar field $\psi$ are taken to obey Dirichlet boundary conditions, i.e. fixing the boundary-induced metric $h_{ij}(r_{B},x)\equiv h^{B}_{ij}(x)$; and bulk fields $\psi(r_{B},x)\equiv\psi_{B}(x)$. The standard dictionary is recovered in the limit $r_{B}\to\infty$. As mentioned in \cite{Hartman:2018tkw}, Dirichlet boundary conditions might be problematic in higher dimensions in the sense of a well-posed initial valued problem \cite{Anderson_2008,An:2021fcq}; while the QFT is not guaranteed to exist; however, \cite{Hartman:2018tkw} provided evidence in favor of (\ref{eq:Dirchdic}), which we treat as our working assumption.

Following this line of work, we consider a sharp radial cutoff in global AdS$_{d+1}$ space, with a geometry (\ref{eq:asymptotic AdS metric}) for $N(r)=r^2+1$, where we can identify the coordinate change with respect to (\ref{eq: AdS with global dS foliation}) for our choice $\ell_{d+1}=1$, $H=1$,
\begin{equation}\label{eq:coordinate change r sigma}
r=\sinh\sigma~.
\end{equation}
Meanwhile, the $T^2$ deformed CFT$_d$ theory corresponding to an AdS$_{d+1}$ bulk dual in Einstein gravity, is described by the action
\begin{equation}
\pdv{I}{\lambda}=\int\rmd^d x\sqrt{h} \qty[\qty(T_{ij}+\frac{\alpha_{d}}{\lambda^{\frac{d-2}{d}}}G_{ij})^2-\frac{1}{d-1}\qty(T_{i}^i+\frac{\alpha_{d}}{\lambda^{\frac{d-2}{d}}}G_i^i)^2]~,
\end{equation}
where $\lambda$ is the $T^2$ deformation parameter, $G_{ij}$ is the Einstein tensor on the surface where the deformed CFT is located, and $\alpha_d$ is a constant which only scales with $G_{d+1}$\footnote{Explicitly:
\begin{equation}
\alpha_d=\begin{cases}
0~,&d=2\\
\frac{1}{8\pi G_{d+1}(d-2)}\qty(\frac{4\pi G_{d+1}}{d})^{1/d}~,&d=3,~4\\
(d^{1-\frac{2}{d}}(d-2)(\pi G_{d+1})^{\frac{2}{d}}2^{1+\frac{4}{d}})^{-1}~,&d>4~.
\end{cases}
\end{equation}}

Using the coordinate change (\ref{eq:coordinate change r sigma}), we can identify the holographic dictionary relating the finite radial cutoff of AdS$_{d+1}$ at $r=\sinh \sigma_c$, and the $T^2$ deformation parameter $\lambda$ as 
\begin{equation}\label{eq:holo dictionary}
\lambda=\frac{4\pi G}{d \qty(\sinh\sigma_c)^d}~.
\end{equation}

\subsection{Multiverse models from the (JT) dS wedge holography}\label{sec:multiverse}
\begin{figure}[t!]
    \centering
    \includegraphics[width=0.9\textwidth]{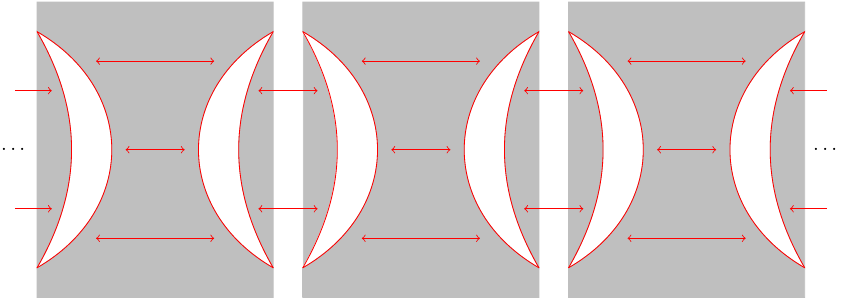}
    \caption{Penrose diagram of the AdS$_3$ space (in gray) cutoff by UV and IR dS regions (in red). We prepare the state with the Hartle-Hawking construction, considering a thermal density matrix for each brane $\rho_{\rm TFD}$ purified by a maximally entangled partner (UV or IR region respectively) with an empty bulk in between (white) and glue them together (red arrows), as shown in the figure. Although the UV and IR regions within a single universe are initially unentangled, as the branes accelerate and exchange Hawking modes.}
    \label{fig:repManifold}
\end{figure}

We now describe a multiverse model by gluing n-copies of dS wedge universes. The different UV and IR regions are pairwise glued together in the configuration illustrated in Fig. \ref{fig:repManifold}.

Double holography provides three descriptions of the system:
\begin{itemize}
    \item[a.] $n$ pairs of Euclidean CFT$_{d-1}$ defects timelike separated from each pair.
    \item[b.] $n$ double-sided IR and $n$ UV dS$_{d}$ regions with a matter theory entangled with each other.
    \item [c.] $n$ double-sided AdS$_{d+1}$ bulk space cutoff in the UV and IR region that glued along pairs of UV/UV or IR/IR regions periodically.
\end{itemize}
In the simplest case, $d=2$, gravity would be topological. To generate a dynamical evolution with respect to a braneworld observer, we add an intrinsic gravitational theory, namely dS JT gravity, on the IR brane.\footnote{\label{fnt:10}Alternatively, we could have added dS JT gravity on both the UV and IR regions. One can immediately see that adding JT on the UV region would modify the phases for the entanglement in (\ref{eq:length}, \ref{eq:Sbrane}) by the same additive (time-dependent) constant. Since only the relative entropy difference enters in the physical analysis, we do not consider this possibility, although it can be straightforwardly included.} We now analyze the conditions for this configuration to be realized.

\subsubsection*{IR branes}
To set the stage, consider dS JT gravity as intrinsic theory on the IR branes, so that the system is described by
    \begin{equation}\label{eq:Total action with JT}
    \int_{Q_b}\rmd^2 x\sqrt{-h}\mathcal{L}_{\rm intrinsic}=\frac{1}{8\pi G_{3}}\int_{Q_b}\rmd^2 x\sqrt{-h}K+I_\text{JT}
\end{equation}
with the JT action given by
\be
\label{eq:JTaction}
    I_\text{JT} = \frac{1}{16\pi G_{\rm b}}\qty[ \int \rmd^2x \sqrt{-h} \, \Phi_0 \tilde{{R}} + \int \rmd^2x \sqrt{-h} \, \Phi \left( \tilde{{R}} - 2\Lambda_{\rm JT} \right)]~.
\ee
Here, $\Lambda_{\rm JT}$ is the cosmological constant on the brane; $G_{\rm b}$ is Newton's gravitational constant on the brane; $\tilde{{R}}$ the Ricci scalar; $\Phi$ is the dilaton, and $\Phi_0$ a topological term, which we will not consider in the subsequent discussion.

The brane equation of motion translates to the Israel junction conditions \cite{Israel:1966rt}
\begin{equation}\label{eq:Neumann any brane}
    \Delta K_{ij}-h_{ij}\Delta K=8\pi G_3 T_{ij}
\end{equation}
where $T_{ij}$ is the effective stress tensor on the brane; while $\Delta K_{ij}=K_{ij}^{(+)}-K_{ij}^{(-)}=2K_{ij}^{(+)}$ is the extrinsic curvature difference between opposite sides, denoted by $\pm$, in the double-sided $Q_{b}$ brane.

On the other hand, the dilaton equation of motion fixes the background geometry on the ETW brane to be
\begin{equation}
    \tilde{{R}} = 2\Lambda_{\rm JT}~.\label{eq:EOM dil}
\end{equation}
Using the induced metric at the location of either brane, where $\sigma$ is fixed, (\ref{eq:main metric}) with (\ref{eq:EOM dil}) gives
\begin{equation}\label{eq: Rtilde LambdaJT}
    \sinh^2\sigma_b=1/\Lambda_{\rm JT}~,
\end{equation}
which implies that $\Lambda_{\rm JT}$ fixes the brane location; i.e. it plays the role of the brane tension.

One can also find the effective stress tensor in the brane with dS JT gravity by performing the variation of (\ref{eq:JTaction}) with respect to $h_{ij}$, resulting in the relation:
\begin{equation}\label{eq:hij JT EOM}
    -\nabla_i\nabla_j\Phi+h_{ij}\nabla^2\Phi+h_{ij}\Lambda_{\rm JT}\Phi=8\pi G_{\rm b}T_{ij}~.
\end{equation}
Matching (\ref{eq:hij JT EOM}) and (\ref{eq:Neumann any brane}) now gives
\begin{equation}
    \nabla_i\nabla_j\Phi+\Lambda_{\rm JT}\Phi h_{ij}=\frac{G_{\rm b}}{G_{3}}\Delta K_{ij}~,
\end{equation}
with the covariant derivative $\nabla_i$ taken with respect to the induced metric $h_{ij}$ on the brane.

Next, we look for solutions of the form
\begin{equation}
    \Phi=\varphi_0+\varphi~,
\end{equation}
with $\varphi$ being the homogeneous solution, i.e.
\begin{equation}\label{eq:homogeneus Eq dilaton}
    \nabla_i\nabla_j\varphi+\Lambda_{\rm JT}\,h_{ij}\,\varphi=0~;
\end{equation}
while $\varphi_0$ is a constant term, determined as
\begin{equation}
    \varphi_0=\frac{G_{\rm b}}{2\Lambda_{\rm JT}G_{3}}\Delta K~.
\end{equation}
Thus, the junction conditions in ETW branes with dS JT gravity can be absorbed into an overall constant term in the dilaton on the branes being glued together.

One can find solutions for (\ref{eq:homogeneus Eq dilaton}) in global dS$_2$ coordinates (\ref{eq: AdS with global dS foliation})
\begin{align}
    \rmd s^2&=\sec^{2}t(-\rmd t^2+\rmd\alpha^2)~,\\
\label{eq:dilaton Rindler AdS}
    \varphi&=\varphi_1\tan t+\varphi_2\frac{\cos\alpha}{\cos t}+\varphi_0
\end{align}
with $\varphi_1$, $\varphi_2$, $\varphi_0\in\mathbb{R}$ being constants. We notice that for $\varphi_1=0$, (\ref{eq:dilaton Rindler AdS}) reproduces the dilaton in the full-reduction model dS JT gravity \cite{Aalsma:2022swk}. It follows that the thermodynamics on the ETW brane also satisfy the relations found for dS JT gravity \cite{Svesko:2022txo}.

As last remark, note that unlike the original formulation of dS wedge holography in \cite{Aguilar-Gutierrez:2023tic}, since we consider a single ETW brane in the bulk, there is no longer issue about perturbative stability due to fluctuations of the ETW brane location.

\section{Coarse-graining of information in the multiverse}\label{sec:coarsegraing multiverse}
In this section, we study the coarse-grained information that can be collected by a UV observer in the multiverse model in \ref{sec:multiverse} using holographic entanglement entropy and compare it with the island formula in false vacuum decay models in quantum cosmology \cite{Aguilar-Gutierrez:2021bns,Levine:2022wos} and the cosmological central dogma \cite{Shaghoulian:2021cef}.

{As mentioned in Sec. \ref{sec:intro}, we stress that having the UV cutoff very close to the asymptotic boundary allows for suppressing non-local effects on the dual QFT (where the scale of non-locality depends on the deformation parameter, as $\sqrt{\lambda}$ \cite{Guica:2019nzm}). The Ryu-Takayanagi (RT) formula \cite{Ryu:2006bv,Ryu:2006ef} is generically not applicable due to the non-locality of the field theory, but it remains well defined at the perturbative level in the deformation parameter $\lambda$ \cite{Banerjee:2024wtl}, or at the non-perturbative one, considering that the deformation parameter is the smallest length scale in the system \cite{Apolo:2023ckr}. For this reason, we place the UV cutoff very near the conformal boundary, meaning that we keep the deformation parameter $\lambda\ll1$ (i.e. the cutoff location $\sigma_c\gg1$ in (\ref{eq:holo dictionary})), where the theory accepts a quasilocal description, and the holographic entanglement entropy is well-defined \cite{Apolo:2023ckr, Banerjee:2024wtl,Demise:2021cfx,Chen:2018eqk,Donnelly:2018bef}.}

\subsection{Information recovery protocol}
Our protocol, based on \cite{Aguilar-Gutierrez:2023tic}, considers an observer collecting Hawking radiation in a single double-sided cutoff region $Q_c$ with Dirichlet boundary conditions. 
Gravity decouples and the effective theory of the UV region is thus described by a dS QFT with a mass gap \cite{Maldacena:2012xp,Fischler:2013fba}. We work in the framework of IR/UV entanglement, where, as explained in \cite{Balasubramanian:2011wt,Aguilar-Gutierrez:2023tic}, the total Hilbert space factorizes as\footnote{The argument employs a Fock space decomposition, $\mathcal{H}=\otimes_{\vec{p}}\mathcal{H}_{\vec{p}}$ with $\mathcal{H}_{\vec{p}}$ the Hilbert space for momentum modes $\vec{p}$, so that the notion of entanglement between UV and IR degrees of freedom in the dual theory are described at different momentum scales \cite{Balasubramanian:2011wt}.}
\begin{equation}\label{totalHilbert}
    \mathcal{H}=\mathcal{H}_{\text{UV}}\otimes\mathcal{H}_{\text{IR}}\,,
\end{equation}
where $\mathcal{H}_{\text{UV}}$ is the Hilbert space of the dS QFT (with a gap) and $\mathcal{H}_{\text{IR}}$ represents the Hilbert space of the IR degrees of freedom, geometrized by $Q_b$. {Also, note that, although there is no gravity localization in the IR brane, this does not alter the evaluation of the HRT formula from the UV boundary.} 

Let $\mathbf{R}$ represent the subregion accessible to the UV observer, such that $\mathbf{R}\subset\partial\hat{\mathcal{M}}$. For simplicity, we will consider an entangling region with disk topology partitioning the $S^{d-1}$ internal space. We thus study a generic entangling region $\mathbf{R}=I\times S^{d-2}$, with $I=\{\alpha\in[\alpha_1,\alpha_2]\}$. However, given the presence of the dS JT couplings in the IR brane, the maximal area surfaces \emph{do not} need to appear at a fixed (global) time (denoted by $\tau$) slice.

In general, if the brane contains an intrinsic gravitational theory, the HRT formula describing the von Neumann entropy with respect to a boundary subregion $\mathbf{R}$ can be expressed by, 
\begin{equation}\label{eq:EE with brane couplings}
    S_{\text{EE}}({\mathbf{R}})=\text{Min}\,\text{Ext}\qty[\frac{A(\Sigma_{\mathbf{R}})}{4G_{d+1}}+\frac{A(\sigma_{\mathbf{R}}=\Sigma_{\mathbf{R}}\cap\text{brane})}{4G_{\rm b}}]
\end{equation}
where $\Sigma_{\mathbf{R}}$ is the bulk HRT surface (homologous to $\mathbf{R}$), and $\sigma_{\mathbf{R}}$ is its intersection with the brane. The interpretation from the brane perspective is seen as a version of the ``island" rule \cite{Geng:2020qvw,Geng:2020fxl,Geng:2021iyq,Chen:2020uac,Chen:2020hmv,Grimaldi:2022suv}.

We now proceed with the evaluation in (\ref{eq:EE with brane couplings}) within a single {$T^2$} dS wedge holographic {universe}, meaning a single AdS$_{3}$ space capped off by the UV and IR regions. We evaluate the HRT surfaces anchored at the finite UV region, whose cutoff surface $\sigma_c\rightarrow\infty$ corresponds to $\rho_c\rightarrow\infty$.

Our interest is to dress the dS braneworlds with dS JT gravity to model false vacuum eternal inflation as in \cite{Aguilar-Gutierrez:2021bns} (a periodic dS$_2$ spacetime with multiple inflationary and black hole patches). We have the option to either add intrinsic gravity on the UV and/or IR regions. In the first case, given that the UV region represents a region with Dirichlet boundary conditions, this choice would only amount to a shift in the entropy (\ref{eq:result integration}) both before and after the Page transition. Since we are only interested in the entropy difference between the two phases, which is unaffected by adding JT on the UV region, we will only include JT on the IR brane, which has a nontrivial modification in the variational problem.

We then need to search for minimal-length surfaces, corresponding to the functional
\begin{equation}\label{eq:length}
\begin{aligned}
    S^{\rm bef\,PT}=&\frac{1}{4G_{3}}\int_{{\rho}_t}^{{\rho}_c}\sqrt{1-\cosh^2\rho\,\tau'({\rho})^2+\sinh^2{\rho}\,\alpha'({\rho})}~\rmd{\rho}~,
\end{aligned}
\end{equation}
where $\rho_t$ is the turning point, in which $\rho'(\alpha)=0$.

Meanwhile, when the RT surfaces land on the IR brane, we consider the entropy functional (where we absorb the constant $\Phi_0$ in $\varphi_0$)
\begin{equation}\label{eq:Sbrane}
\begin{aligned}
    S^{\rm aft\,PT}=&\frac{1}{4G_{3}}\int_{{\rho}_b}^{{\rho}_c}\sqrt{1-\cosh^2\rho\,\tau'({\rho})^2+\sinh^2{\rho}\,\alpha'({\rho})}~\rmd{\rho}+\frac{1}{2G_{\rm b}}\qty[\varphi_2\frac{\cos\alpha_{\rm b}}{\cos t_{\rm b}}+\varphi_0]~.
\end{aligned}
\end{equation}
Here the overall factor of $2$ in the contact comes from the two interceptions of the RT on the branes (see Fig. \ref{fig:Muti braneworlds}).

\subsection{Before the Page transition}
We proceed to determine the conserved charges from (\ref{eq:length}, \ref{eq:Sbrane}) related to the angular momentum and energy in the ambient AdS$_3$ space
\begin{equation}\label{eq:charges}
        E_\alpha=\pdv{\mathcal{L}}{\alpha'}~,\quad E_\tau=\pdv{\mathcal{L}}{\tau'}~,
    \end{equation}
which allows to solve $\alpha(\rho)$ and $\tau(\rho)$. Since the configuration is empty AdS space and the $Q_b$ brane does not play a role before the Page transition, the minimal area surfaces must exist within $\tau=$ constant slices.

For the boundary condition of $\alpha(\rho)$, we take a fixed subregion
\begin{equation}\label{eq:bdy cond 2+1 d}
\alpha(\rho\to\infty)-\alpha_0=\Delta\alpha_c~,    
\end{equation}
from some reference location $\alpha_0$. The solution becomes
\begin{equation}\label{eq:alpha rhog}
    \alpha(\rho)-\alpha_0=\arctan{\frac{\sqrt{2}\tan\Delta\alpha_c\cosh{\rho}}{\sqrt{\cosh{2\rho}-\sec^2\Delta\alpha_c}}}~,
\end{equation}
and we can derive that $E_\alpha=\tan\Delta \alpha_c$. This result allows for the explicit evaluation of (\ref{eq:length}) as
\begin{equation}\label{eq:Lag d=3}
    S^{\rm bef\,PT}=\frac{1}{4G_3}\int_{\rho_t}^{\rho_c}\sqrt{1+\alpha'(\rho)^2\sinh^2{\rho}}~\rmd\rho~,\\
\end{equation}
where there is implicitly (asymptotic boundary) time dependence via the coordinate map between AdS$_{d+1}$ with a dS$_d$ foliation and global AdS$_{d+1}$, (\ref{eq:map global to dS foliation}), given by
\begin{equation}\label{eq:cutoff rho global sigma dS}
\sinh{\rho_c}=\sinh{\sigma_c}\cosh{t_c}~,   
\end{equation}
which grows monotonically with $t_c$ as the global dS$_2$ time with respect to the UV observer (which then grows unbounded). From (\ref{eq:bdy cond 2+1 d}) and (\ref{eq:cutoff rho global sigma dS}), we find
\begin{equation}
    \rho_{t}=\text{arccosh}(\sec(\Delta\alpha_c))\,,
\end{equation}
where $\Delta\alpha_c\in\qty[-\pi/2,\,\pi/2]$. Then, (\ref{eq:Lag d=3}) is reduces to
\begin{equation}
    \begin{aligned}\label{eq:Length d=3}
    S^{\rm bef\,PT}&=\frac{1}{4G_3}\log\abs{\frac{\sqrt{\sinh^2\sigma_c\cosh^2t_c+1}+\sqrt{\sinh^2\sigma_c\cosh^2t_c-\tan^2\Delta\alpha_c}}{\sec\Delta\alpha_c}}~,
    \end{aligned}
\end{equation}
where we employed the map in (\ref{eq:map global to dS foliation}). Notice that this thermal entropy grows monotonically with $t_c$ and is unbounded. 

In the following, we show that an observer in a given (double-sided) finite cutoff UV region will detect an increase in the von Neumann entropy until reaching a Page transition.

\subsection{Transition without JT couplings}\label{sec:NO JT}
Given the presence of the IR brane at $\sigma=\sigma_b$, there will be a cut-off scale for the growth in the von Neumann entropy detected by the UV observer. The corresponding entanglement entropy should be evaluated with Neumann boundary conditions on the IR brane, which determines the type of ansatz to be used. Without the JT couplings, one finds a constant entropy from (\ref{eq:Sbrane}), which is expressed in global dS$_2$ coordinates as
\begin{equation}\label{eq:result integration}
    S^{\rm aft\,PT}=\frac{1}{2G_3}(\sigma_c-\sigma_b)~.
\end{equation}
There is an entanglement phase transition between the island (\ref{eq:Sbrane}) and the disconnected phase, which is determined by both $\sigma_b$ and $\alpha(\sigma_c)$. To see that, notice that the disconnected phase in (\ref{eq:Lag d=3}) has a minimum at $t=0$, and that when $\sigma_c\gg1$, we get
\begin{equation}
    S^{\rm bef\,PT}=\frac{1}{4 G_3}\qty(\sigma_c-\log\abs{\sec\Delta\alpha})~.
\end{equation}
Then, comparing with (\ref{eq:Sbrane}), we require
\begin{equation}
\sigma_b<\log\abs{\sec\Delta\alpha}    \label{eq:cond4transition}
\end{equation}
in order to have a transition. Thus we have an avatar of the picture in \cite{Cohen:1998zx} relating the UV and IR cutoffs of the effective field theory (EFT) (i.e. the multiverse braneworld model in our case) from the entropy transition.\footnote{We thank Dominik Neuenfeld for discussions on this point.}

One can straight-forwardly include boundary time dependence in the previous argument at late times, where $\rho_{c}\simeq\sigma_{c}+t_c$ from (\ref{eq:map global to dS foliation}). In that case, one finds a Page time at
\begin{equation}
    t_{P}=\log\abs{\sec\Delta\alpha}-\sigma_b~.
\end{equation}
The result indicates that the time to produce a transition is enhanced by increasing the size of the region where radiation is collected, and there is a threshold given by the location of the IR brane. We then require $\Delta\alpha\simeq\frac{\pi}{2}$ for the late time assumption to be valid.

\subsection{De Sitter JT gravity on the IR brane} \label{sec:JT IR brane}
We will now perform the extremization in (\ref{eq:Sbrane}) again, but now the conserved charge $E_\alpha$ no longer vanishes at the brane location. We start with the global AdS coordinates (\ref{eq:global coordinates}), with two conserved charges in $d=2$. One can then use the variation of the total action and the solutions to the EOM to evaluate the charge with Neumann boundary conditions. We find,
\begin{equation}\label{eq:Sbrane final}
\begin{aligned}
    S^{\rm aft\,PT}=\frac{1}{2G_{\rm b}}\qty[\varphi_{1}\tan t_b+\varphi_{2}\frac{\cos\alpha_b}{\cos t_b}+\varphi_0]+\frac{1}{16 G_3}\log\frac{(1+\Delta(\sigma_c,\,t_c))(1-\Delta(\sigma_b,\,t_b))}{(1-\Delta(\sigma_c,\,t_c))(1+\Delta(\sigma_b,\,t_b))}~,
\end{aligned}
\end{equation}
where
\begin{equation}\label{eq:Delta sigma t}
    \Delta(\sigma,\,t)=\frac{E_\tau^2-E_\alpha^2+1+2\sinh^2\sigma\cosh^2 t}{2
   \sqrt{\left(E_\tau^2-E_\alpha^2+1\right) \sinh ^2\sigma\cosh^2 t-E_\alpha^2+\sinh ^4\sigma\cosh^4 t}}.
\end{equation}
Next, we would like to find out the boundary conditions for the system with JT couplings. For that, notice:
\begin{equation}
\begin{aligned}
    \delta S^{\rm aft\,PT}=&\int_{{\rho}_b}^{{\rho}_c}\rmd 
    s\qty(\partial_{\alpha}\mathcal{L}-\dot{E}_{\alpha}+\partial_{\tau}\mathcal{L}-\dot{E}_{\tau})+\eval{E_\alpha\delta\alpha}_{{\rho}_b}^{{\rho}_c}+\eval{E_{\tau}\delta{\tau}}_{{\rho}_b}^{{\rho}_c}\\
    &+\frac{1}{2G_{\rm b}}\qty[\varphi_{1}\sec^2t_b\delta t_b+\varphi_{2}\frac{\sin\alpha_b}{\cos t_b}\delta\alpha_b+\varphi_{2}\frac{\cos\alpha_b\sin t_b}{\cos^2t_b}\delta t_b]~.
\end{aligned}
\end{equation}
Imposing Dirichlet boundary conditions at the finite UV cutoff region, the vanishing of the variation above gives us
\begin{equation}
\delta\alpha({\rho}_c)=0,\quad \delta\tau({\rho}_c)=0   ~.
\end{equation}
Using the map (\ref{eq:map global to dS foliation}) and considering that the location of the branes is fixed (i.e. $\delta\sigma_{\rm brane}=0$); we can deduce the boundary conditions at the location of the IR brane as:
\begin{align}
    &\delta t_b:\quad E_\tau=\frac{1}{2G_{\rm b}}\frac{\tanh^2\sigma_b \sinh ^2t_b+1}{\tanh \sigma_b \cosh t_b}\qty[\varphi_{1}\sec^2t_b+\varphi_{2}\frac{\cos\alpha_b\sin t_b}{\cos^2t_b}],\label{eq:Etau}\\
    &\delta \alpha_b:\quad E_\alpha=\frac{\varphi_{2}}{2G_{\rm b}}\frac{\sin\alpha_b}{\cos t_b}~.\label{eq:Ealpha}
\end{align}
We may now evaluate the functional $S^{\rm aft\,PT}$ in (\ref{eq:Sbrane final}) subject to (\ref{eq:Etau}, \ref{eq:Ealpha}). However, we need to express $t_b$ in terms of the physical parameters measured by the UV observer, namely the angular distance for collecting the radiation $\alpha(\sigma_c)$, and the physical time $t_{c}$. We will then have to solve the equations of motion of the extremal area surfaces and impose boundary conditions at $\sigma_c$, to determine those that minimize $S^{\rm after~ PT}$.

Solving the equations of motion resulting from (\ref{eq:Sbrane}) in terms of the conserved charges $E_\tau$ and $E_\alpha$ in (\ref{eq:charges}) leads to the solution
\begin{align}
    \tan \left(2 \left(\alpha (\rho )-\alpha _0\right)\right)&=\frac{\left(1-E_{\alpha }^2+E_{\tau }^2\right) \sinh ^2(\rho )-2 E_{\alpha }^2}{2 E_{\alpha }
   \sqrt{\left(1-E_{\alpha }^2+E_{\tau }^2\right) \sinh ^2(\rho)-E_{\alpha }^2+\sinh ^4(\rho )}}\label{eq:alpha rho}\\
   \tan\qty(2(\tau(\rho)-\tau_0))&=\frac{\left(E_{\tau }^2-E_{\alpha }^2-1\right) \cosh^2\rho-2 E_{\tau }^2}{2
   E_{\tau } \sqrt{\left(E_{\tau }^2-E_{\alpha }^2-1\right)\cosh^2\rho-E_\tau^2+\cosh^4\rho}}\label{eq:tau rho}
\end{align}
where $\alpha_0$ and $\tau_0$ are an arbitrary angular reference point and an arbitrary reference time at which we start measuring the Hawking radiation, respectively.

We may use the transformation to dS global coordinates (\ref{eq: AdS with global dS foliation}) to express (\ref{eq:alpha rho}) as 
\begin{equation}\label{eq:Solving sigma}
    \tan \left(2 \left(\alpha (\sigma )-\alpha _0\right)\right)=\frac{\left(-E_{\alpha }^2+E_{\tau }^2+1\right) \sinh ^2(\sigma ) \cosh ^2(t)-2 E_{\alpha }^2}{2 E_{\alpha }
   \sqrt{\left(-E_{\alpha }^2+E_{\tau }^2+1\right) \sinh ^2(\sigma ) \cosh ^2(t )-E_{\alpha }^2+\sinh ^4(\sigma ) \cosh ^4(t )}}~.
\end{equation}
Evaluating (\ref{eq:Solving sigma}) at $\sigma_{\rm c}\rightarrow\infty$, we find a relation between the angle measured by the UV observer and the conserved charges:
\begin{equation}\label{eq: Delta alpha constraint}
    \boxed{2 E_{\alpha } \tan (2 \Delta \alpha)=- E_{\alpha }^2+E_{\tau }^2+1}
\end{equation}
where $\Delta \alpha=\alpha_c-\alpha_0$. 

Next, we simplify (\ref{eq:tau rho}) with the dS global coordinates (\ref{eq: AdS with global dS foliation}):
\begin{equation}\label{eq:tau(alpha)}
\begin{aligned}
    &\tanh(2\Delta\tau)=\frac{\left(E_{\tau }^2-E_{\alpha }^2-1\right) \left({\sinh ^2\sigma \cosh ^2t}+1\right)-2 E_{\tau }^2}{2
   E_{\tau } \sqrt{\left(E_{\tau}^2-E_{\alpha}^2-1\right)\left({\sinh ^2\sigma \cosh ^2t}+1\right)-E_{\tau }^2+\left({\sinh ^2\sigma \cosh ^2t}+1\right)^2}}~.
\end{aligned}
\end{equation}
where $\Delta\tau=\tau(\rho)-\tau_0$. At the location of the UV region, we get
\begin{equation}\label{eq:tau c}
    \boxed{2E_\tau\tanh(2\Delta\tau_c)=E_{\tau }^2-E_{\alpha }^2-1~.}
\end{equation}
Let us now solve (\ref{eq: Delta alpha constraint}, \ref{eq:tau c}) to determine two branches of solutions for $E_\alpha$ and $E_\tau$:
\begin{align}
    E^{(\pm)}_\tau=\tan2\Delta\tau_c\frac{\sec^22\Delta\alpha_c\pm{\sec2\Delta\alpha_c\sec2\Delta\tau_c}}{\tan^22\Delta\alpha_c-\tan^22\Delta\tau_c}~,\label{eq:Etau c}\\    E_\alpha^{(\pm)}=\frac{\tan2\Delta\alpha_c
    \sec^22\Delta\tau_c
    \pm{\sec2\Delta\alpha_c
    \sec2\Delta\tau_c}}{\tan^22\Delta\alpha_c-\tan^22\Delta\tau_c}\label{eq:Ealpha c}~.
\end{align}
Moreover, we have a relation between the charges above with the brane coordinates $t_b$, $\alpha_b$ in (\ref{eq:Etau}, \ref{eq:Ealpha}), which is enough to determine the solution to the extremization problem.

To solve the above relations we will consider perturbative solutions in $\alpha_b\ll1$, $t_b\ll 1$, while keeping $\sigma_b$ arbitrary. In that case, combining (\ref{eq:Etau}, \ref{eq:Ealpha}) 
and taking $\varphi_1=0$\footnote{This choice is required for the dS JT gravity dilaton to have SdS asymptotics \cite{Aguilar-Gutierrez:2021bns}.} gives:
\begin{align}
    &t_b=\frac{2G_{\rm b}}{\varphi_{2}}\tanh \sigma_bE_\tau+\mathcal{O}\qty(\frac{G_b^2}{\varphi_2^2})~,\label{eq:sol tb}\\
    &\alpha_b= \frac{2G_{\rm b}}{\varphi_{2}}E_\alpha+\mathcal{O}\qty(\frac{G_{\rm b}^2}{\varphi_{2}^2})~.\label{eq:sol alphab}
\end{align}
Notice that this approximation becomes increasingly better as $\frac{\varphi_2}{G_b}\ll 1$, which corresponds to the semiclassical regime of dS JT gravity.

\begin{figure}[t!]
    \centering
    \includegraphics[width=0.6\textwidth]{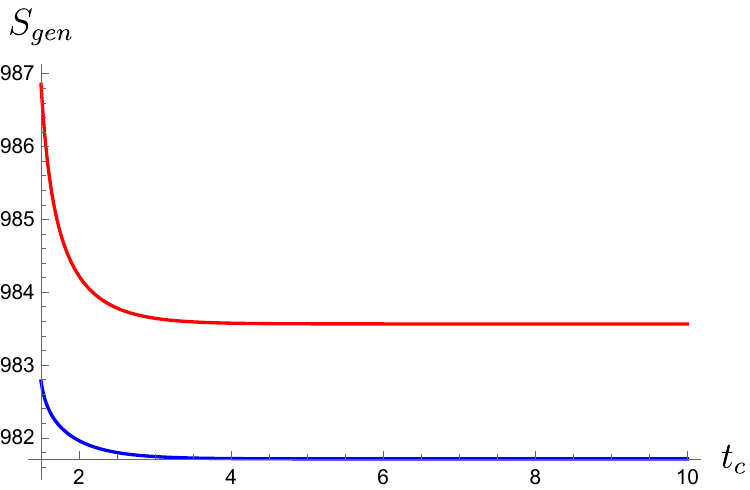}
    \caption{Comparison between $S^{\rm aft~PT}$ (\ref{eq:S after full}) for the different root of the conserved charges $E_\tau^{(\pm)}$, $E_\alpha^{(\pm)}$ in (\ref{eq:Etau c}) and (\ref{eq:Ealpha c}) respectively. We have used $\Delta\alpha_c=\frac{\pi}{6}$, $t_{c,~0}=0$, $\epsilon_c=1$, $G_3=1/16$, $\sigma_b=5$, $G_{\rm b}=1/2$, $\varphi_0=1$, $\varphi_1=0$ and $\varphi_2=1000$. Notice that the generalized entropy for the $E_{\tau,\,\alpha}^{(-)}$ roots (in blue) dominates over the $E_{\tau,\,\alpha}^{(+)}$ roots (red).}
    \label{fig:Isl saddles}
\end{figure}

Using the trigonometric identity
\begin{equation}
    \tan\qty(2(\tau-\tau_0))=2\frac{\tan \tau\left(1-\tan ^2\tau_0\right)-\tan \tau_0 \left(1-\tan ^2\tau\right)}{{(1-\tan ^2\tau)(1-\tan ^2\tau_0)}+{4 \tan \tau_0 \tan \tau}}~,
\end{equation}
and the map to global dS space coordinates (\ref{eq:map global to dS foliation}), we can express $\Delta\tau_c$ purely in terms of $t_c$ and $t_{c,\,0}$ (the reference time for the Page curve, at which the UV observer starts collecting the Hawking radiation):
\begin{equation}\label{eq:Tau c tc tc0}
    \tan2\Delta\tau_c=2\frac{\sinh t_c\left(1-\sinh ^2t_{c,\,0}\right)-\sinh t_{c,\,0} \left(1-\sinh ^2t_{c}\right)}{{(1-\sinh ^2t_{c})(1-\sinh ^2t_{c,\,0})}+{4 \sinh t_{c,\,0} \sinh t_{c}}}~.
\end{equation}
With all tools at hand, we can evaluate the island transition using (\ref{eq:Sbrane final}, \ref{eq:Delta sigma t}) as
\begin{equation}\label{eq:S after full}
\begin{aligned}
    S^{\rm aft\,PT}=\frac{1}{2G_{\rm b}}\qty[\varphi_{2}\frac{\cos\alpha_b}{\cos t_b}+\varphi_0]+\frac{1}{16 G_3}\log\abs{\frac{2}{\epsilon_c}\frac{1-\Delta(\sigma_b,\,t_b)}{1+\Delta(\sigma_b,\,t_b)}}~,
\end{aligned}
\end{equation}
where $\epsilon_c$ is a UV regulator, and we have set $\varphi_1=0$. Since there are two roots in (\ref{eq:Etau c}) and (\ref{eq:Ealpha c}), we must explore the one that produces the minimum entropy to identify the island contribution. The comparison between the roots is displayed in Fig. \ref{fig:Isl saddles}, which shows that the dominating saddle is the one determined through $E_\tau^{(-)}$, $E_\alpha^{(-)}$ in (\ref{eq:Etau c}, \ref{eq:Ealpha c}). Moreover, one can notice that the entropy for both curves starts decreasing until it reaches an asymptotic late time value, which we deduce below. The decrease in von Neumann entropy can then be used for information recovery using protocols such as \cite{Aalsma:2022swk}.

We now deduce the late-time expression for the entropy with the above relations. Using the ansatz (\ref{eq:sol tb}, \ref{eq:sol alphab}), one finds
\begin{equation}\label{eq: Safter PT late}
    S^{\rm aft~PT}=\frac{1}{2G_b}(\varphi_2+\varphi_0)+\frac{1}{16G_3}\log\abs{\frac{2}{\epsilon_c}\frac{1-\Delta_b}{1+\Delta_b}}~.
\end{equation}
In the limit $t_c\gg1$, (\ref{eq:Tau c tc tc0}) gives
\begin{equation}
    \tan2\Delta\tau_c=\frac{2\sinh t_{c,~0}}{1-\sinh^2 t_{c,~0}}~,
\end{equation}
while when $t_{c,~0}=0$ and $\tau_c\rightarrow\infty$, we have $\Delta\tau_c\rightarrow0$. Therefore, from (\ref{eq:Delta sigma t}), one recovers
\begin{equation}
    \Delta(t_b,\,\sigma_b)\simeq\frac{E_\tau^2-E_\alpha^2+1+2\sinh^2\sigma_b}{2
   \sqrt{\left(E_\tau^2-E_\alpha^2+1\right) \sinh ^2\sigma_b-E_\alpha^2+\sinh ^4\sigma_b}}
\end{equation}
and therefore, the von Neumann entropy collected by a UV observer at $t_c\gg1$ for generic $\Delta \alpha_b$ is just the constant given in (\ref{eq: Safter PT late}).

\begin{figure}[t!]
    \centering
    \includegraphics[width=0.65\textwidth]{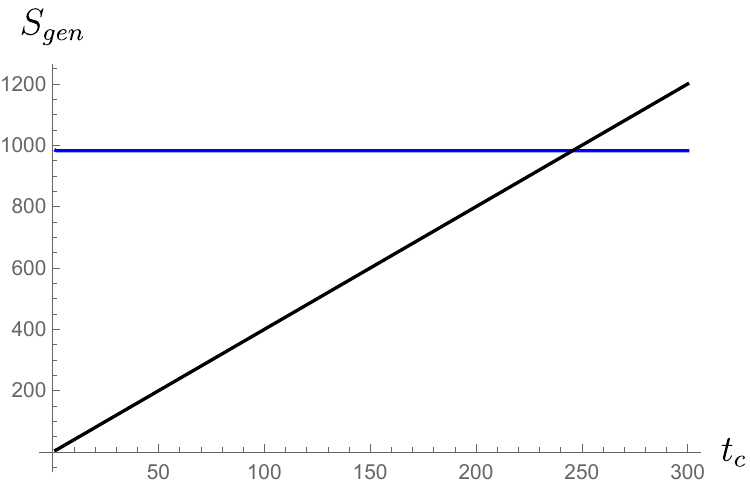}
    \caption{Page curve in a braneworld model with intrinsic dS JT gravity couplings, where the black curve indicates the Hawking phase $S^{\rm bef~PT}$ (\ref{eq:Length d=3}), and the blue curve $S^{\rm aft~PT}$ (\ref{eq:S after full}), with the same parameters as Fig. \ref{fig:Isl saddles}.}
    \label{fig:single KR Universe}
\end{figure}

Meanwhile, the late-time asymptotics of the Hawking entropy (\ref{eq:Length d=3}) can be deduced as
\begin{equation}
    S^{\rm bef~PT}\simeq\frac{1}{4G_3}\log\abs{\frac{2}{\epsilon_{c'}}\cos\Delta\alpha_c\cosh t_c}\sim \frac{t_c}{4G_3}~,
\end{equation}
where $\epsilon_{c'}$ is a UV regulator when $\sigma_c\rightarrow\infty$, and in the last expression we have discarded $t_c$ independent additive terms. The plot of the Page curve, displaying the dominating island and the Hawking phase is shown in Fig. \ref{fig:single KR Universe}.\footnote{The resulting island phase differs from other approaches in the literature to model multiverses from Karch-Randall branes \cite{Yadav:2023qfg}, where instead, it was suggested that the total entropy should be the sum of the individual contribution from each brane and its bath. The difference with respect to our model is that the UV and IR regions intercept only at two boundary times (see Fig. \ref{fig:UV IR branes}), while the latter approach patches them all together in a codimension-two conformal defect.}

\subsection{Central dogma and quantum cosmology interpretation}\label{sec:central dogma}
\begin{figure}[t!]
    \centering
    \includegraphics[width=0.8\textwidth]{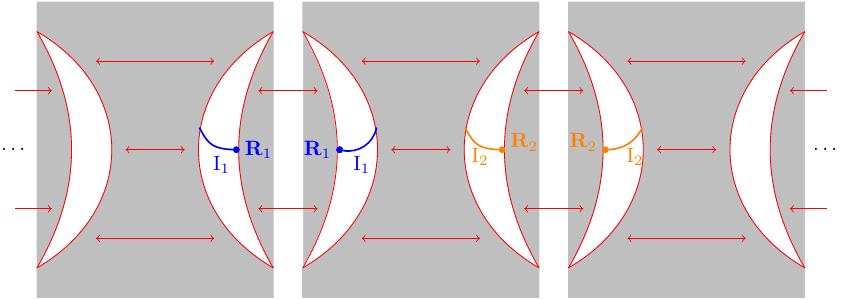}\vspace{0.5cm}\\
    \includegraphics[width=0.8\textwidth]{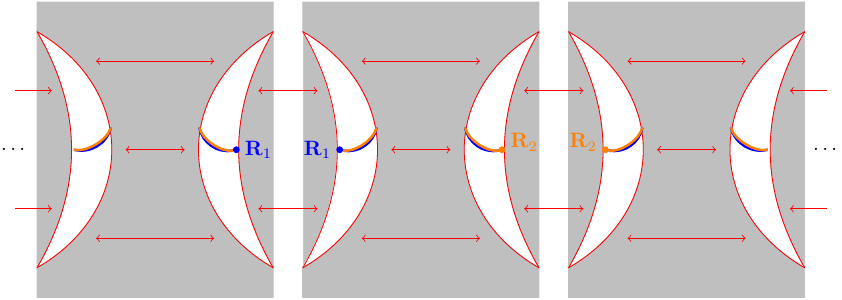}
    \caption{Island saddles for the multiverse model, where $\mathbf{R}_1$ (blue dot) and $\mathbf{R}_2$ (orange dot) represent different codimension-one bath regions in the finite cutoff UV at a given global time $\tau_c$ when the observers are collecting information in each brane universe. \textit{
    Top}: Islands ($\mathbf{I}_1$, $\mathbf{I}_2$, blue and orange solid curves with the same color scheme as the corresponding bath regions) connecting a single side of the UV region to an IR brane, and producing the dominant contribution to the entropy (\ref{eq:S after full}). \textit{Bottom}: A different possibility where $I_1$ and $I_2$ always overlap.}
    \label{fig:Muti braneworlds}
\end{figure}

The previous calculation shows a non-trivial island between the UV and IR region from the perspective of a single { observer}. A natural possibility is to have an island that crosses multiple patches, as shown in Fig. \ref{fig:Muti braneworlds} (bottom). In that case, the von Neumann entropy is the $NS^{\rm aft~PT}$ found above, with $N$ the number of patches. However, the dominating saddle is the one where the island only intercepts the IR brane at a single time, see Fig \ref{fig:Muti braneworlds} (top), which is given by (\ref{eq:S after full}). In contrast, the island saddle encompassing different universes would allow for overlapping HRT surfaces once we introduce another observer on a different double-sided UV region, shown also in Fig. \ref{fig:Muti braneworlds} (bottom). Then, the theory might be in tension with bulk reconstruction, as there would be causally disconnected UV observers sharing a common entanglement wedge, corresponding to the bulk AdS space, as this would modify the commutation relations between the observables accessible to the different UV regions. { This could lead to a violation of the no-cloning theorem in quantum mechanics as two observers could reconstruct the same information simultaneously.}

A similar paradox in the context of dS JT gravity and the connection with the central dogma for cosmological horizons and the no-cloning theorem was discussed in \cite{Levine:2022wos}, as we mentioned in the introduction.
We can view the island transitions in our model through the lenses of quantum cosmology, which shares several similarities to \cite{Aguilar-Gutierrez:2021bns,Levine:2022wos} \footnote{A technical difference with these other models is that our multiverse configuration 
does not have conical singularities.}. 
The information available to a single universe where gravity is non-dynamical (i.e. the finite UV cutoff region) and includes the observers at other universes {due to the gluing conditions requiring that all the universes are completely symmetric.}. The microscopic degrees of freedom can be probed by observers on the UV cutoff region. Yet, the observers can only probe the regions available through the islands due to the coarse-graining of information in dS wedge holography, which prohibits communication between observers at different branes.

Thus, the proposed extension of dS wedge holography provides a conceptually new insight into this problem in the sense that the coarse-graining of information in quantum cosmology, at least within our setting, involves entanglement between the UV and IR energy modes. It would be of interest to see whether this observation occurs in an explicit higher dimensional model of false vacuum decay in eternal inflation, such as \cite{Garriga:2006hw}.

\section{Discussion}\label{sec:discussion}
In this work, we have constructed a multiverse toy model based on the dS wedge holography proposal in \cite{Aguilar-Gutierrez:2023tic}, and $T^2$ deformations \cite{Hartman:2018tkw}. The multiverse arises by generating several copies of global AdS$_3$ space with a finite radial cutoff in the UV arising from turning a $T^2$ deformation in the dual theory, and a dS$_2$ ETW brane cutting off the interior, which are glued together in a periodic matter through the Israel junction conditions. To have an interesting evolution that resembles false vacuum eternal inflation, while keeping the model exactly solvable, we ``dressed" the IR brane as a near Nariai black hole using the dS JT gravity as the intrinsic gravity theory on it. 
In principle, we could add JT couplings to the finite UV cutoff region as well (after performing the dual deformation), but as we showed this would not qualitatively change our study in any way.

We studied the coarse-graining of local measurements expected in false vacuum eternal inflation \cite{Aguilar-Gutierrez:2021bns} within our toy model, by evaluating the fine-grained von Neumann entropy (and islands) with the HRT formula. Double holography allowed us to determine the entanglement entropy of Hawking radiation collected by an observer in the UV cutoff, which we associated with this coarse-graining.

The presence of dS JT gravity on the IR brane modifies the Page curve transition and allows for information recovery with respect to a given dS braneworld observer. Interestingly, we find an avatar of a previous proposal \cite{Cohen:1998zx}, where a generic effective field theory in curved spacetime is expected to involve a relation between UV and IR cutoffs, in order {to satisfy} consistency relations in the effective field theory description. With our model, one arrives at a similar conclusion, as the Page time depends on both UV and IR quantities. {Furthermore, the interplay between UV and IR degrees of freedom is clear from the the RT surfaces connecting these regions, giving a non-vanishing entanglement entropy; the
existence of the IR brane is what allows the appearance of the connected RT surfaces and the Page transition.} Finally, the resulting entanglement of Hawking modes captured by a UV cutoff observer also shows agreement with previous arguments about the central dogma for cosmological horizons and consistency with the no-cloning theorem \cite{Levine:2022wos}. Namely, there are no overlapping island saddles when we allow for multiple Dirichlet brane observers.

There are some obvious questions to be addressed in the (near) future.

Firstly, most of our work on entanglement in the multiverse model has been focused on the AdS$_{3}$ bulk perspective, and the experience of the UV observers. However, it would be interesting to have an interpretation of the coarse-graining in quantum cosmology from the time-like separated codimension-two defects on the interception between the branes (see the brane interceptions at $\tau=\pm\frac{\pi}{2}$ in Fig. \ref{fig:UV IR branes}), as this might provide a dS/CFT holographic perspective \cite{Strominger:2001pn} for the coarse-graining of the UV and IR degrees of freedom.

Secondly, to study the coarse-graining of information in quantum cosmology we have focused on evaluating the von Neumann entropy of spacetime subregions; however, the case with a single universe previously studied in \cite{Aguilar-Gutierrez:2023tic} also considered the evolution of holographic complexity using the C=Volume \cite{Susskind:2014rva,Stanford:2014jda} and C=Anything \cite{Belin:2021bga, Belin:2022xmt} proposals in codimension-one slices, where it was observed that the hyperfast growth of complexity previously found in pure dS space also occurs in dS wedge holography. Moreover, it was recently studied in \cite{Aguilar-Gutierrez:2024rka} that holographic complexity in asymptotically dS spacetimes with multiple inflating and black hole patches can lead to a drastic modification in the evolution of different codimension-one and codimension-zero proposals. Given that the setting considered in \cite{Aguilar-Gutierrez:2024rka} is supposed to be a toy model of eternal inflation as ours, it would be interesting to study the universality of their observations within our setting, to see if the complexity observables also capture a large redundancy of information, as we have argued in our work.

Thirdly, there have been recent discussions about causality violations due to faster-than-light communication in AdS braneworld models with an effective theory in the IR brane \cite{Neuenfeld:2023svs}, where it has been shown that the apparent violations in the braneworld EFT are not visible above its cutoff length scale. It is unclear how these arguments would be modified for the dS braneworlds, in particular in the presence of (dS) JT couplings. Studying the candidate regions replacing the domain of dependence proposed in \cite{Neuenfeld:2023svs} for our model could confirm that the EFT description of Karch-Randal models is consistent with causality in double holography.

Lastly, adding an observer in dS space \cite{Chandrasekaran:2022cip,Witten:2023xze,Witten:2023qsv,Aguilar-Gutierrez:2023odp,Jensen:2023yxy,Kudler-Flam:2023qfl,Faulkner:2024gst} has led to many developments regarding algebraic studies physical observables and to define generalized entropies. It might be useful to rigorously investigate the algebra of observables within our model to provide a better understanding of the different entropy transitions found in our work and their connection with coarse-graining in quantum cosmology.

\section*{Acknowledgements}
We thank Thomas Hertog, Ayan K. Patra, Juan Pedraza, Marika Taylor, and Edgard Shaghoulian for useful discussions; and especially Dominik Neuenfeld for early collaboration and helpful comments. SEAG thanks the University of Amsterdam, the Delta Institute for Theoretical Physics, and the International Centre for Theoretical Physics for their hospitality and financial support during several phases of the project, and the Research Foundation - Flanders (FWO) for also providing mobility support (Grant No. K250423N). The work of SEAG is partially supported by the FWO Research Project G0H9318N and the inter-university project iBOF/21/084. FL is grateful for the hospitality of Nordita Institute for Theoretical Physics where parts of this work were carried out. The work of FL is supported by STFC(ST/W507799/1). 

\appendix

\bibliographystyle{JHEP}
\bibliography{references.bib}
\end{document}